\documentclass{aa}
\usepackage{txfonts}
\usepackage{graphicx}
\usepackage{subcaption}
\usepackage[colorlinks=true,allcolors=blue]{hyperref}

\begin{document} 

\titlerunning{}
\authorrunning{Mountrichas et al.}
\titlerunning{SF rates and histories of AGN and non AGN across different density fields}

\title{Probing star formation rates and histories in AGN and non-AGN galaxies across diverse cosmic environments and extensive X-ray luminosity ranges}

\author{G. Mountrichas\inst{1}, M. Siudek \inst{2} and O. Cucciati \inst{3}}
          
     \institute {Instituto de Fisica de Cantabria (CSIC-Universidad de Cantabria), Avenida de los Castros, 39005 Santander, Spain
              \email{gmountrichas@gmail.com}
              \and
              Institute of Space Sciences (ICE, CSIC), Campus UAB, Carrerde Can Magrans, s/n, 08193 Barcelona, Spain
              \and
              INAF - Osservatorio di Astrofisica e Scienza dello Spazio di Bologna,
via Gobetti 93/3, 40129 Bologna, Italy
              }

\abstract{In this work, we compare the star-formation rates (SFRs) and star-formation histories (SFHs) of AGN and non-AGN galaxies. We explore these aspects across different density fields and over three orders of magnitude in X-ray luminosity (L$_X$). For that purpose, we employ X-ray AGN detected in the XMM-XXL field and construct a galaxy control sample, using sources from the VIPERS catalogue. We apply strict photometric and quality selection criteria to ensure that only sources with robust (host) galaxy measurements are included in the analysis. Our final samples consist of 149 X-ray AGN with $\rm 42<log,[L_{X,2-10keV}(ergs^{-1})]<45$ and 3\,488 non-AGN systems. The sources span a redshift range of $\rm 0.5<z<1.0$  and have stellar masses within, $10.5<\rm log\,[M_*(M_\odot)]<11.5$. For these systems, there are available measurements for their local densities and their spectral lines (D$_n$4000) from the VIPERS catalogue. To compare the SFR of these two populations, we calculate the SFR$_{norm}$ parameter. The latter is defined as the ratio of the SFR of AGN to the SFR of non-AGN galaxies with similar M$_*$ and redshift. Our findings reveal that low and moderate L$_X$ AGN ($\rm 42<log,[L_{X,2-10keV}(ergs^{-1})]<44$) that live in low density fields have a nearly flat SFR$_{norm}-$L$_X$ relation. In contrast, AGN of similar L$_X$ that live in high density environments present an increase of SFR$_{norm}$ with L$_X$. These results are in line with previous studies. Notably, our results suggest that the most luminous of the AGN ($\rm log,[L_{X,2-10keV}(ergs^{-1})]>44$) exhibit increased SFR relative to non-AGN galaxies, and this trend appears to be independent of the density of the environment. Furthermore, for AGN with similar L$_X$, those in high-density regions tend to have higher SFR$_{norm}$ values compared to their counterparts in low-density areas. Comparison of the D$_n$4000 spectral index, which serves as a proxy for the age of the stellar population, reveals that low-to-moderate L$_X$ AGN live in galaxies with comparable stellar populations with non-AGN systems, regardless of the density field they live in. However, the most luminous X-ray sources tend to live in galaxies that have younger stellar populations than non-AGN galaxies, regardless of the galaxy's environment.}

   
\maketitle

\section{Introduction}

In the last couple of decades, there has been significant advancement in our comprehension of the formation and growth of supermassive black holes (SMBHs) throughout cosmic history. Black holes increase in size through the accretion of cold gas onto their accretion disks. This gas originates from regions at least nine orders of magnitude larger in scale, either within the host galaxy or from the broader extragalactic environment \citep[e.g.,][]{Alexander2012}. During this accretion phase, the SMBH becomes active, transforming the galaxy into an active galactic nucleus (AGN). Simultaneously, star formation is driven by cold gas. Notably, AGN activity and star formation tend to reach their peaks during the same cosmic era (at around $\rm z \sim 2$) \citep[e.g.,][]{Boyle1998, Sobral2013, Madau2014}. These alignments strongly suggest a connection between SMBH activity and the growth of galaxies. 

To explore this connection, numerous previous studies have examined the potential correlation between X-ray luminosity (L$_X$), which serves as a proxy for the accretion power of SMBHs, and the star formation rate (SFR) within the host galaxy. While not all research findings concur, the overall consensus suggests that the relationship between SFR and L$_X$ is notably impacted by the specific range of L$_X$ and stellar mass (M$_*$) that is under consideration \citep[e.g.,][]{Mountrichas2024a, Cristello2024}.

In particular, results from recent studies seem to suggest that in galaxies with $10.5<\rm log\,[M_*(M_\odot)]<11.5$ that host luminous AGN ($\rm log,[L_{X,2-10keV}(ergs^{-1})]>44$), the SFR is elevated compared to galaxies without AGN. On the contrary, galaxies that host low-to-moderate L$_X$ AGN ($\rm 42<log,[L_{X,2-10keV}(ergs^{-1})]<44$) appear to have lower or at most equal SFR to non-AGN galaxies, with similar M$_*$ and redshift \citep{Mountrichas2021c, Mountrichas2022a, Mountrichas2022b}. Notably, recent work by \cite{Mountrichas2024a} extends these findings to both lower ($\rm log\,[M_*(M_\odot)]<10.5$) and higher ($\rm log\,[M_*(M_\odot)]>11.5$) stellar mass regimes. However, the threshold L$_X$ for observing this elevated SFR appears to increase as we transition from less to more massive galaxies.

These observations hint at the possibility that distinct physical mechanisms may drive AGN activity in varying mass and/or luminosity regimes. AGN, as a diverse population, are subject to different triggering processes, as indicated by both observational studies \citep[e.g.,][]{Allevato2011, Mountrichas2013, Mountrichas2016} and theoretical models \citep[e.g.,][]{Fanidakis2012, Fanidakis2013}. Various AGN triggering mechanisms may dominate depending on factors such as redshift, L$_X$, and the mass of the source. One plausible mechanism linking star formation with AGN activity involves galaxy mergers \citep[e.g.,][]{Hopkins2008a}. This scenario posits that a fraction of the cold gas available for star formation in galaxies accretes onto the SMBH, activating AGN. In higher-mass systems, alternative mechanisms decoupled from the host galaxy's star formation may fuel the SMBH. For instance, the SMBH might be activated when diffuse hot gas in quasi-hydrostatic equilibrium accretes onto the SMBH without first cooling onto the galactic disc. Semi-analytic models suggest that this mechanism becomes dominant in very massive systems \citep[e.g.,][]{Fanidakis2013}. Importantly, this framework implies that any observed correlation between AGN activity and SFR is a consequence of a shared mechanism influencing both properties. In other words, there may not be an inherent, direct connection between the two parameters.

Earlier investigations that examined the correlation between SFR and L$_X$ \citep[e.g.,][]{Masoura2018, Bernhard2019, Florez2020, Mountrichas2022a}, have not delved into the influence of the cosmic environment on the SFR-L$_X$ relationship. In the case of galaxies (i.e., non-AGN systems) this aspect has been the focus of extensive research both in the local universe and at higher redshifts \citep[e.g.,][]{Elbaz2007, Scoville2013, Lin2014, Erfanianfar2016, Cooke2023}. However, these investigations typically do not consider the activity of the SMBH, meaning they do not distinguish between systems with AGN and those without.

In a recent study, \cite{Mountrichas2023c} compared the SFR of AGN and non-AGN galaxies, in different density fields (and for different morphologies). Their research utilized sources in the COSMOS field, that probed a redshift range of $\rm 0.3<z<1.2$ and their M$_*$ fell within $10.5<\rm log\,[M_*(M_\odot)]<11.5$. Their results showed that AGN in low density environments have lower SFR compared to non-AGN systems. Conversely, in higher density fields, the SFR of AGN became equivalent or even surpassed that of non-AGN galaxies, particularly in cases with elevated L$_X$ ($\rm 43<log\,[L_{X,2-10keV}(ergs^{-1})]<44$). Furthermore, by employing two spectral indices which are commonly used as indicators of stellar population ages, namely the D$_n$4000 and H$_\delta$, the study revealed that low L$_X$ AGNs ($\rm log\,[L_{X,2-10keV}(ergs^{-1})]<43$) have consistent stellar populations in all cosmic environments, while moderate L$_X$ AGNs ($\rm 43<log\,[L_{X,2-10keV}(ergs^{-1})]<44$) tend to have younger stars and are more likely to have undergone a recent burst in high-density fields. Non-AGN galaxies, on the other hand, tended to exhibit older stellar populations and were less prone to having undergone recent episodes of star formation in denser environments when compared to their isolated counterparts. Moreover, based on their analysis, bulge-dominated (BD) AGN were scarse in denser environments, while in the case of non-AGN galaxies, the fraction of BD systems increases with the density field. 

These findings may suggest that in dense environments, AGN feedback counteracts gas removal, preventing the quenching of star formation—a phenomenon known as positive feedback \citep[e.g.,][]{Zinn2013, Santoro2016, Meenakshi2022}. The more active the AGN (higher L$_X$), the more effective the AGN feedback becomes. However, as mentioned earlier, an alternative interpretation could be that a shared mechanism, such as galaxy mergers fuels both star formation and the SMBH, triggering both processes simultaneously. This perspective aligns with the observed variations in morphological types between AGN and non-AGN systems in different environments, hinting at galaxy interactions as triggers for both phenomena. Moreover, the scarcity of BD galaxies in AGN-hosting systems within dense fields may indicate interactions that prompt both star formation and AGN activity. Finally, the lower SFRs observed in galaxies hosting AGNs, compared to non-AGN systems in low-density fields, could be attributed to a limited availability of gas in these environments. For example, \cite{Zubovas2013} demonstrated that in galaxies undergoing a gas-poor phase, AGN feedback might quench the star formation process. It is important to note, however, that a limitation of the \cite{Mountrichas2023c}  study was the absence of the most luminous AGNs, specifically those with X-ray luminosities exceeding $\rm log\,[L_{X,2-10keV}(ergs^{-1})]>44$.

In this study, we employ X-ray AGN detected in the XMM-XXL field and galaxies from the VIPERS catalogue, to compare the SFR and star-formation histories (SFH) of the two populations, in different density fields and across three orders of magnitude in L$_X$. Our goal is to leverage the broader range of high X-ray luminosities probed by the AGN included in the XMM-XXL dataset, thus expanding the scope of these inquiries to encompass higher X-ray luminosity levels compared to the work of \cite{Mountrichas2023c}. The structure of the paper is as follows. In Sect. \ref{sec_data}, we provide an overview of the data we utilize for our investigations. In Sect. \ref{sec_analysis}, we describe the spectral energy distribution (SED) fitting analysis we perform and outline how we obtain information on the SFH and the local densities of the sources. We also describe the various photometric and quality selection criteria we apply. The results of our analysis are presented in Sect. \ref{sec_results}. Finally, Sect. \ref{sec_conclusions} offers a summary of our key findings.
 
\section{Data}
\label{sec_data}
In our study, we utilize two distinct samples. The first dataset comprises X-ray-detected AGN, while the second one incorporates data from optical spectra measurements for over 90,000 galaxies. Below, we present details about these two datasets.

\subsection{The X-ray dataset}
\label{sec_xray_data}

The X-ray sample consists of X-ray AGN detected in the northern field of the {\it{XMM-Newton}}-XXL survey \citep[XMM-XXL;][]{Pierre2016}. XMM-XXL is a medium-depth X-ray survey that covers a total area of 50\,deg$^2$ split into two fields nearly equal in size: the XMM-XXL North (XXL-N) and the XXM-XXL South (XXL-S). Furthermore, it is the largest XMM-Newton project approved to date ($>6$\,Ms) with median exposure at 10.4\,ks and a depth
of $\rm \sim 6 \times 10^{-15}\,erg\,s^{-1}\,cm^{-2}$ for point sources, in the 0.5-2\,keV band. In this study, we use the XXL-N sample that consists of 14\,168 sources. To identify the X-ray detections at other wavelengths the X-ray counterparts have been cross-matched with optical, NIR, and MIR surveys \citep[for more details see][]{Chiappetti2018}.

\subsection{The galaxy control sample}
\label{sec_vipers_data}

The galaxy control sample used in this work comes from the public data release 2 \citep[PDR-2;][]{Scodeggio2016} of the VIPERS survey \citep{Guzzo2014, Garilli2014}, which partially overlaps with the XMM-XXL field. The observations were carried out using the VIMOS \citep[VIsible MultiObject Spectrograph,][]{LeFevre2003} on the ESO Very Large Telescope (VLT). The survey covers an area of $\approx$ 23.5\,deg$^2$, split over two regions within the  Canada-France- Hawaii Telescope Legacy Survey (CFHTLS-Wide) W1 and W4 fields. Follow-up spectroscopic targets were selected to the magnitude limit i$^\prime=22.5$  from the T0006 data release of the CFHTLS catalogues. An optical colour-colour pre-selection, namely, {\it{[(r-i)$ > $0.5(u-g) or (r-i)$ > $0.7]}}, excludes galaxies at $z<0.5$, yielding a $>98\%$ completeness for $z>0.5$ and up to $\rm z\sim 1.2$ \citep[for more details see][]{Guzzo2014}. Then, PDR-2 consists of 86,775 galaxies with available spectra. Each spectrum is assigned a quality flag that quantifies the redshift reliability. In all the VIPERS papers, redshifts with flags in the range between 2 and 9 have been considered to be reliable and are those used in the science analysis \citep{Garilli2014, Scodeggio2016}. The above criteria yield 45,180 galaxies within the redshift range spanned by the VIPERS survey (0.5$<$z$<$1.2). 

The process of adding multiwavength photometry to the galaxy sample is described in detail in Sect. 2.1 in \cite{Mountrichas2019} \cite[see also][]{Mountrichas2023d}. In brief, the VIPERS galaxy catalogue was cross-matched with sources in the VISTA Hemisphere Survey \citep[VHS,][]{McMahon2013} and the AllWISE catalogue from the WISE survey \citep{Wright2010}, using the xmatch tool from the astromatch\footnote{https://github.com/ruizca/astromatch} package. The process is described in detail in Sect. 2.5 in \cite{Pouliasis2020}. xmatch matches a set of catalogues and gives the Bayesian probabilities of the associations or non-association \citep{Pineau2017}. We only kept sources with a high probability of association ($>68\%$). When one source was associated with several counterparts, we selected the association with the highest probability. 14,128 galaxies from the VIPERS catalogue have counterparts in the near- (VISTA) and mid- (WISE) infrared parts of the electromagnetic spectrum.

\section{Galaxy properties}
\label{sec_analysis}

In this section, we describe how we obtain information about the properties of the sources used in our analysis. Specifically, we present how we measure the SFR and M$_*$ of AGN and non-AGN galaxies and how we retrieve knowledge on their stellar populations.

\subsection{Calculation of SFR and M$_*$}
\label{sec_sed}

The (host) galaxy properties of both the X-ray AGN and the galaxies in the control sample have been calculated via SED fitting, using the CIGALE code \citep{Boquien2019, Yang2020, Yang2022}. The SED fitting analysis is described in detail, for instance, in Section 3.1 in \cite{Mountrichas2021b, Mountrichas2022a, Mountrichas2022b, Mountrichas2023c}. In brief, the galaxy component is modelled using a delayed SFH model with a function form $\rm SFR\propto t \times exp(-t/\tau)$. A star formation burst is included \citep{Ciesla2017, Malek2018, Buat2019} as a constant ongoing period of star formation of 50\,Myr. Stellar emission is modelled using the single stellar population templates of \cite{Bruzual_Charlot2003} and is attenuated following the \cite{Charlot_Fall_2000} attenuation law. To model the nebular emission, CIGALE adopts the nebular templates based on \cite{Inoue2011}. The emission of the dust heated by stars is modelled based on \cite{Dale2014}, without any AGN contribution. The AGN emission is included using the SKIRTOR models of \cite{Stalevski2012, Stalevski2016}. CIGALE has the ability to model the X-ray emission of galaxies. In the SED fitting process, the observed L$_X$ in the $2-10$\,keV band are used, provided by the \cite{Marchesi2016}. The parameter space used is shown in Table 1 in \cite{Mountrichas2022a}. The reliability of the SFR measurements, both in the case of AGN and non-AGN systems, has been examined in detail in our previous works and, in particular, in Sect. 3.2.2 in \cite{Mountrichas2022a}. For instance, using CIGALE's ability to create and analyse mock catalogues based on the best-fit model of each source of the dataset, the effect of the (lack of) far-infrared (Herschel) and ultraviolet photometry has been investigated, among others. Furthermore, consistent SFR measurements have been obtained via SED fitting analysis and those derived from the OII $\rm \lambda 3726$ line \citep[e.g.,][]{Siudek2023a}.

\subsection{Star-formation history}
\label{sec_sfh}

The VIPERS catalog offers insights into the stellar populations of the galaxies it encompasses, focusing on the measurement of the D$_n$4000 spectral index. D$_n$4000 tends to be low for younger stellar populations and high for older, metal-rich galaxies, as demonstrated by previous studies \citep[e.g.,][]{Kauffmann2003a}. The narrow definition of the D$_n$4000 spectral indicator presented in \cite{Balogh1999} has been adopted \citep{Siudek2017}. The equivalent width of the H$_\delta$ spectral line also provides important information for the stellar population of a galaxy, as it rises rapidly in the first few hundred million years after a burst of star formation, when O- and B-type stars dominate the spectrum, and then decreases when A-type stars fade \citep[e.g.,][]{Kauffmann2003a, Wu2018}. However, H$_\delta$ is particularly sensitive to the spectral resolution of the spectra and due to the resolution of VIPERS spectra could not be measured \citep{Siudek2017}. Therefore, in this study we make use of the available D$_n$4000 measurements of the VIPERS catalogue to examine the stellar populations of the AGN and non-AGN systems included in our datasets.

\subsection{Local densities}
\label{sec_DF}

The main goal of this work is to study the star-formation rates and histories of AGN and non-AGN galaxies, across different density fields. The VIPERS galaxy catalogue includes information on the local densities of the sources. Their calculation is described in \cite{Cucciati2014, Cucciati2017, Siudek2022, Siudek2023b}. In brief, the local environment is characterized by the density contrast, $\delta$, smoothed over a cylinder with a radius equal to the distance of the fifth nearest neighbour \citep{Cucciati2017}, taking into account the local density and the mean density at each redshift. The mean density is estimated using all galaxies (both spectroscopic and photometric) that trace the density field within a cylindrical top-hat filter. The tracers are selected either by applying a cut $\rm M_B \leq (20.4-z)$, that provides complete tracer samples up to $\rm z<0.9$ or are extended out to $\rm z=1$ using even brighter tracers. In our analysis, we use overdensities calculated with tracers up $\rm z=1$, since it increases our sample size by $\sim 10\%$. However, we confirm that our results and overall conclusions are not sensitive to this choice. 

\begin{figure}
\centering
  \includegraphics[width=0.9\linewidth, height=7cm]{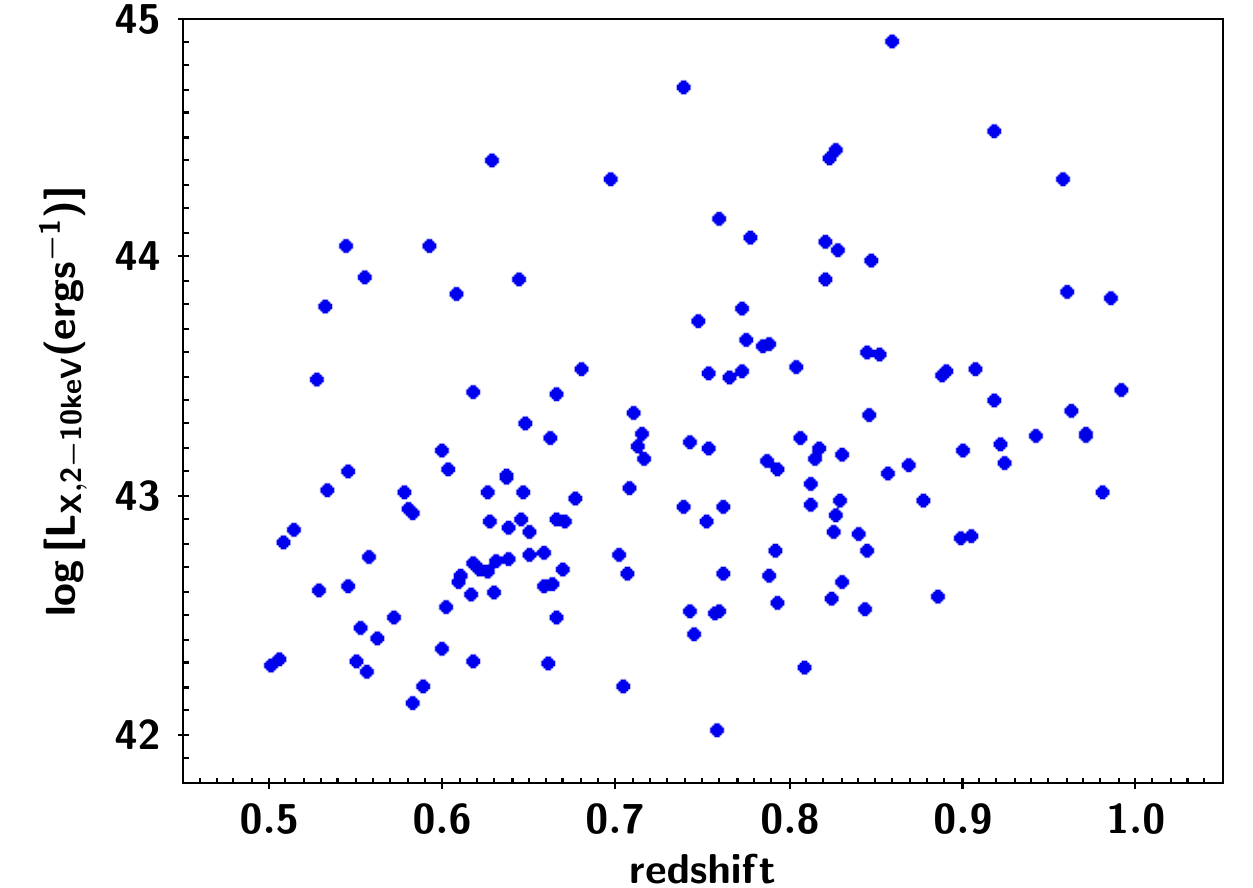}
  \caption{The distribution of the 149 X-ray AGN used in our analysis, in the L$_X-$redshift plane.}
  \label{fig_lx_redz}
\end{figure} 

\begin{figure}
\centering
  \includegraphics[width=0.9\linewidth, height=7cm]{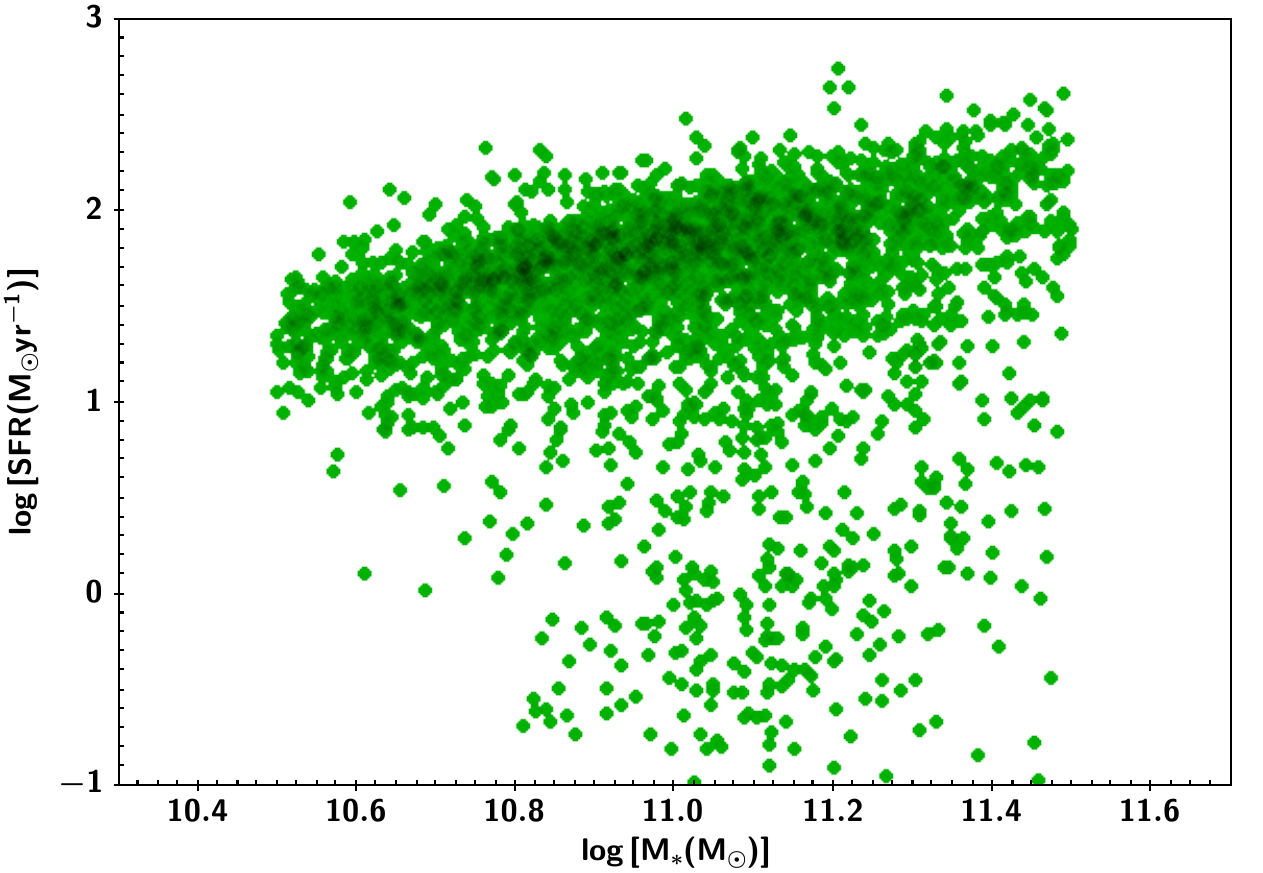}
  \caption{The distribution of galaxies in the control sample, in the SFR-M$_*$ plane.}
  \label{fig_sfr_mstar_vipers}
\end{figure} 

\subsection{Selection criteria and final samples}
\label{sec_final_samples}

In this section, we outline the criteria we apply for assembling the
final dataset of X-ray sources, drawn from the XMM-XXL catalogue (\ref{sec_xray_data}) and the final control sample of non-AGN galaxies, extracted from the VIPERS survey (\ref{sec_vipers_data}).

\subsubsection{The final X-ray sample}
\label{sec_final_xray}

To gain information about the environments that the X-ray AGN live in, we cross-match the XXM-XXL dataset with the VIPERS sample. This results in 940 sources. In our analysis, we need to use only systems with the most reliable M$_*$ and SFR calculations. For that purpose, we require AGN to have measurements in optical, near-infrared (J, H, K) and mid-infrared (W1, W2, W4) part of the spectrum (see Sect. \ref{sec_xray_data}). This criterion is met by 555 X-ray sources. Furthermore, to exclude systems with bad SED fits and unreliable host galaxy measurements, we impose a reduced $\chi ^2$ threshold of $\chi _r^2<5$ \citep[e.g.][]{Masoura2018, Buat2021}. We also exclude sources for which CIGALE could not constrain the parameters of interest (SFR, M$_*$), by following the criteria applied in prior studies \citep[e.g.,][]{Mountrichas2021b, Koutoulidis2022, Garofalo2022, Pouliasis2022, Mountrichas2023a, Mountrichas2023b, Mountrichas2023d}. Specifically, we utilize the two values that CIGALE estimates for each calculated galaxy property. One value corresponds to the best model and the other value (bayes) is the likelihood-weighted mean value. A large difference between the two calculations suggests a complex likelihood distribution and important uncertainties. We therefore only include in our analysis sources with $\rm \frac{1}{5}\leq \frac{SFR_{best}}{SFR_{bayes}} \leq 5$ and $\rm \frac{1}{5}\leq \frac{M_{*, best}}{M_{*, bayes}} \leq 5$, where SFR$\rm _{best}$ and  M$\rm _{*, best}$ are the best-fit values of SFR and M$_*$, respectively and SFR$\rm _{bayes}$ and M$\rm _{*, bayes}$ are the Bayesian values estimated by CIGALE. There are 455 AGN that fulfill these requirements. Out of these 455 X-ray sources, 174 have information about their density field (\textsc{density$>-90$}) and meet the following criterion, {\textsc{density\_mask$ \geq 0.6$}}. The latter requirement reassures that at least $60\%$ of the cylinder volume overlaps with the VIPERS survey footprint \citep[for more details see][]{Davidzon2016, Cucciati2017}. 

Prior investigations have emphasized the significance of M$_*$ in the comparative analysis of SFR between galaxies hosting AGN and those without \citep[e.g.,][]{Mountrichas2021c, Mountrichas2022a, Mountrichas2022c, Mountrichas2024a}. Specifically, these earlier studies suggested that the L$_X$ threshold at which AGN-hosting systems exhibit elevated SFR increases with the mass of the galaxy citep{Mountrichas2024a}. This threshold appears to differ across different stellar mass ranges, namely $\rm log\,[M(M_\odot)]<10.5$, $\rm 10.5<log\,[M_*(M_\odot)]<11.5$, and $\rm log\,[M_*(M_\odot)]>11.5$. Consequently, we confine our X-ray dataset to galaxies within the range $\rm 10.5<log\,[M_*(M_\odot)]<11.5$, resulting in a selection of 149 AGN meeting this criterion. These X-ray sources cover a redshift span of $\rm 0.5<z<1.0$ and exhibit X-ray luminosities within $\rm 42<log\,[L_{X,2-10keV}(ergs^{-1})]<45$, as illustrated in Fig. \ref{fig_lx_redz}. It is noteworthy that within this specified M$_*$ range, both our X-ray dataset and the galaxy control sample achieve completeness in mass \citep{Davidzon2013, Mountrichas2023a}.

\subsubsection{The final galaxy control sample}
\label{sec_final_vipers}

We apply the same photometric, quality and M$_*$ criteria, described in Sect. \ref{sec_final_xray}, on the galaxy control sample. Furthermore, we exclude X-ray sources that are included in the XMM-XXL catalogue and systems with a significant AGN component, revealed by the SED fitting process. Specifically, we use the CIGALE measurements and exclude sources that have AGN fraction, $\rm frac_{AGN}>0.2$ \citep[e.g.,][]{Mountrichas2021c, Mountrichas2022a, Mountrichas2022b}. The $\rm frac_{AGN}$ is defined as the fraction of the total IR emission originating from the AGN. This criterion excludes $\sim 60\%$ of the sources in the galaxy control sample, in line with prior studies \citep[e.g.,][]{Georgakakis2017, Salerno2019, Mountrichas2022b}. There are 3\,488 galaxies that meet these requirements. Their distribution in the SFR-M$_*$ plane is presented in Fig. \ref{fig_sfr_mstar_vipers}.

\section{Results}
\label{sec_results}

In this section, we examine how the SFR of AGN and non-AGN galaxies differs as a function of the L$_X$, across different cosmic environments. Furthermore, we investigate the SFH of these two populations for different density fields. 

\begin{figure}
\centering
  \includegraphics[width=0.9\linewidth, height=7cm]{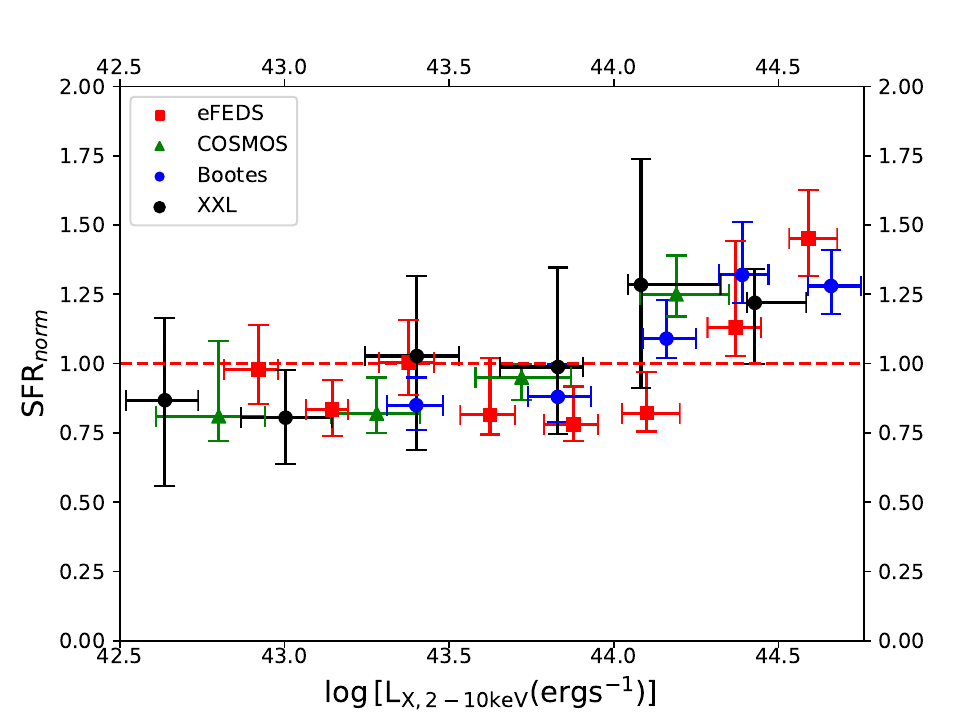}
  \includegraphics[width=0.9\linewidth, height=7cm]{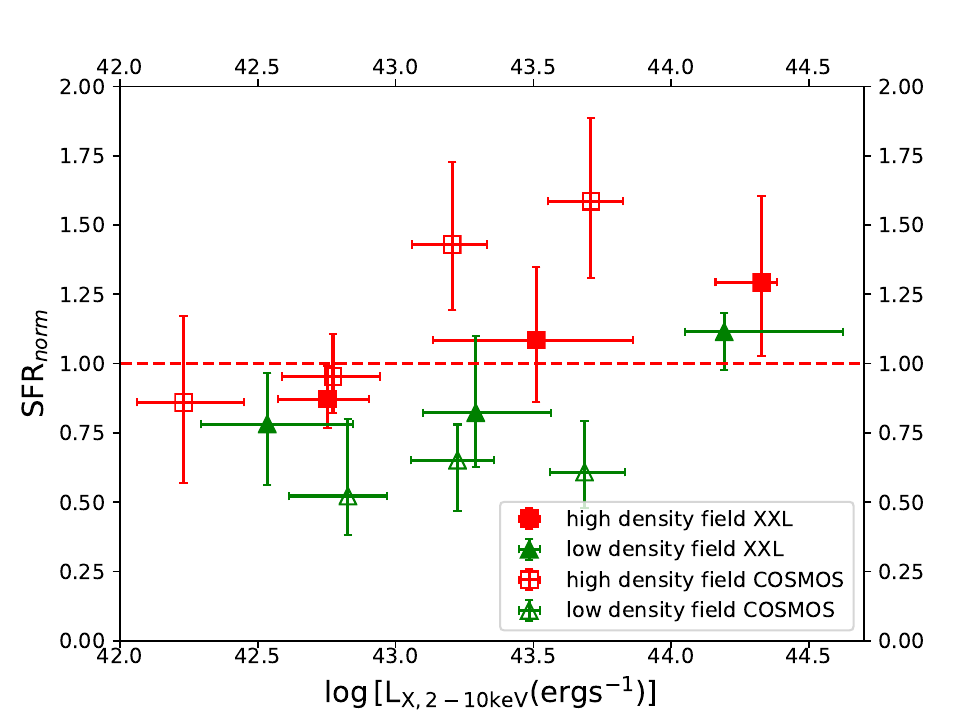}
  \caption{SFR$_{norm}$ ($\rm =\frac{SFR_{AGN}}{SFR_{non-AGN}}$) as a function of L$_X$. Median values are presented. The errors presented are $1\,\sigma$, calculated using bootstrap resampling. The top panel presents the SFR$_{norm}-$L$_X$ relation for the total AGN and galaxy control samples (i.e., irrespective of the cosmic environment). For comparison, results from previous studies are also included. The bottom panel presents the SFR$_{norm}-$L$_X$ relation for sources in high and low density fields (see text for more details). The results from \cite{Mountrichas2023c} using data in the COSMOS field (open symbols) have been incorporated for comparison. }
  \label{fig_sfrnorm_lx}
\end{figure}

\subsection{Comparison of the SFR of AGN and non-AGN galaxies, as a function of L$_X$ and cosmic environment}

To conduct a comparative analysis of the SFR of AGN and non-AGN galaxies, we utilize the SFR$_{norm}$ parameter \citep[e.g.,][]{Mullaney2015, Masoura2018, Bernhard2019}. SFR$_{norm}$ is defined as the ratio of the SFRs of galaxies that host AGN to the SFR of non-AGN systems with similar M$_*$ and redshift. \cite{Mountrichas2021b} showed that systematic biases may affect the measurement of SFR$_{norm}$, when analytical expressions from the literature are employed to estimate the SFR of non-AGN galaxies \citep[e.g.][]{Schreiber2015}, subsequently impacting the accuracy of SFR$_{norm}$ measurements. To mitigate these systematic effects it is preferable to use a galaxy control sample of non-AGN sources. Therefore, to measure SFR$_{norm}$, we use the X-ray and galaxy control samples described in Sect. \ref{sec_data} and follow the process described in previous studies \citep[e.g.][]{Mountrichas2021c, Mountrichas2022a, Mountrichas2022b, Mountrichas2023d}. 

Specifically, for the calculation of SFR$_{norm}$, the SFR of each X-ray AGN is divided by the SFR of galaxies in the control sample that are within $\pm 0.2$\,dex in M$_*$, $\rm \pm 0.075\times (1+z)$ in redshift and are in similar density fields. Furthermore, each source is weighted based on the uncertainty of the SFR and M$_*$ measurements made by CIGALE. Then, the median values of these ratios are used as the SFR$_{norm}$ of each X-ray AGN. We note that our measurements are not sensitive to the choice of the box size around the AGN. Selecting smaller boxes, though, has an effect on the errors of the calculations \citep{Mountrichas2021c}.

To classify sources into density fields, we rely on the overdensity values of the galaxy control sample. We designate sources in 'high density fields' as those with overdensity values falling within the top 30\% of the galaxy sample, while sources with overdensity values within the bottom 30\% are classified as 'low density field' sources. Specifically, sources in the high density field group exhibit $\rm log(1+\delta) > 0.57$, while those in the low density field group have $\rm log(1+\delta) < 0.18$. Among these, there are 43 AGN and 1\,047 non-AGN galaxies in low density fields, as well as 46 AGN and 1,047 non-AGN galaxies in high density fields. This underscores the diversity of environments where AGN can be found, consistent with the findings of \cite{Siudek2023a}, who observed this trend for AGN hosted by dwarf galaxies. It is worth noting that our results and conclusions remain unaffected when using the highest or lowest 20-40\% of overdensity values for source classification.

The top panel of Fig. \ref{fig_sfrnorm_lx} displays the SFR$_{norm}$ as a function of L$_X$, using the samples described in the previous section, indicated by the black circles. Median values are presented. The errors presented are $1\,\sigma$, calculated using bootstrap resampling. For comparison, we have also incorporated measurements obtained from datasets within the Bo$\ddot{o}$tes, COSMOS and eFEDS fields, presented in \cite{Mountrichas2021b, Mountrichas2022a, Mountrichas2022b}. Although, the statistical uncertainties associated with our SFR$_{norm}$ measurements are large, our findings closely align with the outcomes observed in these prior studies. Specifically, the SFR of low-to-moderate L$_X$ ($\rm log\,[L_{X,2-10keV}(ergs^{-1})]<44$) AGN appears lower or, at most, equal to the SFR of non-AGN galaxies, whereas more luminous AGN present enhanced SFR compared to non-AGN systems.  It is important to note that in the depicted results of the previous studies in the figure, quiescent systems have been excluded from both the X-ray and galaxy control datasets. This exclusion primarily impacts the magnitude of the SFR$_{norm}$ calculations, causing a reduction, but it does not alter the underlying trends \citep{Mountrichas2021c, Mountrichas2023c}.

The primary objective of this work is to study the SFR$_{norm}-$L$_X$ relation, across different cosmic environments. The results are displayed in the bottom panel of Fig. \ref{fig_sfrnorm_lx}, indicated by filled symbols. We observe that AGN with low-to-moderate luminosities that live in low density fields (filled triangles) exhibit a flat SFR$_{norm}-$L$_X$ relation, with SFR values lower than those  of non-AGN galaxies in similar environments (i.e., SFR$_{norm}<1$). Conversely, AGN in high density fields (filled squares) display an increase in SFR$_{norm}$ with L$_X$. 

At $\rm log,[L_{X,2-10keV}(erg,s^{-1})]>44$, our results imply that SFR$_{norm}$ increases at elevated L$_X$, irrespective of whether the AGN are situated in high or low-density environments. Furthermore, it appears that, in all cosmic fields, luminous AGN exhibit heightened SFR in comparison to non-AGN systems. However, due to the limited size of our X-ray sample, we can calculate only two data points, which come with relatively large associated uncertainties. Consequently, we refrain from drawing strong conclusions. Nevertheless, it is worth noting that these observations find support in the analysis of SFH of (luminous) AGN and non-AGN galaxies, as we will discuss in the following section.

 In the bottom panel of Fig. \ref{fig_sfrnorm_lx}, we also superimpose results from \cite{Mountrichas2023c} that focused on sources in the COSMOS field, represented by open symbols. Our results show similar trends with these previous findings for AGN featuring low-to-moderate L$_X$. We note, that the overdensity values for the sources in the COSMOS field have been calculated, by applying a different method \citep[i.e., 'weighted adaptive kernel smoothing'; for more details see][]{Darvish2015, Yang2018b} compared to that for the VIPERS galaxies  and is applied to fainter galaxies compared to our galaxy dataset. It is essential to acknowledge that these methodological differences may introduce systematic biases when making comparisons between the two studies. However, \citep{Siudek2023b} examined the field densities of red passive galaxies from VIPERS and COSMOS, revealing a general agreement. This implies that the biases introduced by different techniques may not significantly impact the ultimate conclusions.

The findings of \cite{Mountrichas2023c} also suggest a higher SFR$_{norm}$ amplitude for the same L$_X$ in high density fields compared to low density regions. Our findings align with this trend, yet the distinction is less prominent, despite being consistently observed across various L$_X$. Previous studies have found differences in the SFR of AGN host galaxies for different obscuration levels, in particular for AGN with modest L$_X$ \citep{Georgantopoulos2023, Mountrichas2024b, Mountrichas2024c}. The COSMOS field benefits from deeper observations compared to the XMM-XXL, resulting in a higher number of AGN in COSMOS that exhibit elevated obscuration levels. Therefore, part of the variation in SFR$_{norm}$ between high and low density fields in COSMOS could be attributed to the differing fractions of obscured AGN in these distinct cosmic environments.

Our findings are in agreement with previous research indicating fluctuations in the SFR across diverse density fields, particularly evident at low redshifts \citep[z<1; e.g.,][]{Erfanianfar2016, PerezMillan2023, Sobral2022}. However, other works suggest no discernible SFR variation across different environments \citep[][]{Leslie2020, Cooke2023}. Meanwhile, other investigations have observed environmental impacts solely on the fraction of quiescent galaxies, with no apparent influence on star-forming systems \citep[][]{Darvish2016, Delgado2022}.

\begin{figure}
\centering
  \includegraphics[width=0.95\columnwidth, height=7cm]{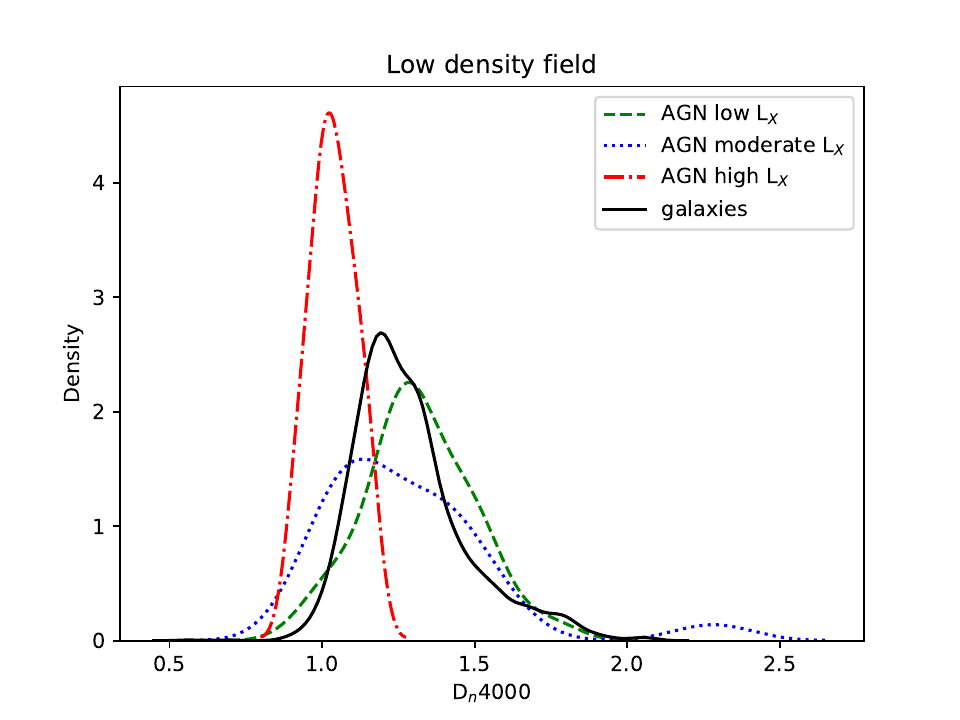}   
  \includegraphics[width=0.95\columnwidth, height=7cm]{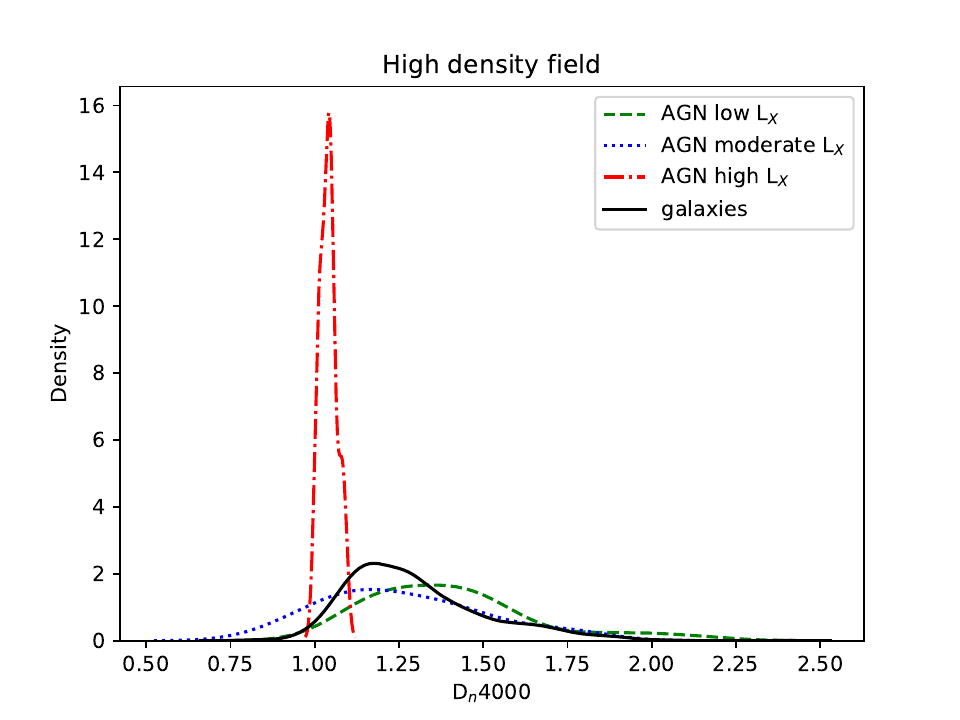} 
  \caption{Distributions of the D$_n$4000 spectral index. The top panel, illustrates the D$_n$4000 distributions for sources in low density fields. The bottom panel, presents the D$_n$4000 distributions for sources in high density fields. For the galaxy control sample, the distributions are shown by the black, solid lines. The distributions of AGN at low ($\rm 42<log\,[L_{X,2-10keV}(ergs^{-1})])<43$), moderate ($\rm 43<log\,[L_{X,2-10keV}(ergs^{-1})])<44$) and high L$_X$ ($\rm 44<log\,[L_{X,2-10keV}(ergs^{-1})])<45$), are presented by the green dashed, blue dotted and red dashed-dotted lines, respectively.}
  \label{fig_sfh}
\end{figure}

\begin{table*}
\caption{Number of X-ray AGN and sources in the control sample, in low and high density fields. Median D$_n$4000 values of each subset is also presented.}
\centering
\setlength{\tabcolsep}{3mm}
\begin{tabular}{ccccccc}
 \hline
 & \multicolumn{3}{c}{low density field} &  \multicolumn{3}{c}{high density field} \\
 \hline
L$_X$ (erg/s) & $10^{42-43}$ & $10^{43-44}$ & $10^{44-45}$ & $10^{42-43}$ & $10^{43-44}$ & $10^{44-45}$ \\
 \hline
no. AGN & 22 & 17 & 4 & 19 & 22 & 5 \\
D$_n4000$\,(AGN) & 1.29 & 1.16 & 1.05 & 1.34 & 1.21 & 1.04  \\ 
no. galaxies & \multicolumn{3}{c}{1047} & \multicolumn{3}{c}{1047}   \\
D$_n4000$ (galaxies) & \multicolumn{3}{c}{1.25} & \multicolumn{3}{c}{1.29} \\
  \hline
\label{table_sfh_numbers}
\end{tabular}
\end{table*}

\subsection{Assessing the star formation histories of AGN and non-AGN galaxies with respect to their L$_X$ and density field}

In this section, we examine the SFH of AGN and non-AGN galaxies across different cosmic environments and for different AGN power. The results of our analysis, are shown in the top panel of Fig. \ref{fig_sfh} and in Table \ref{table_sfh_numbers}. We find that low L$_X$ AGN have a D$_n$4000 distribution that peaks at higher values compared to the D$_n$4000 distribution of non-AGN galaxies. However, applying a Kolmogorov-Smirnov test yields a p$-$value of 0.28, implying that the difference of the two distributions is statistically significant at a level of $<2\,\sigma$ (two distributions differ with a statistical significance of $\sim 2\,\sigma$ for a $\rm p-value$ of 0.05). Moreover, moderate L$_X$ AGN present a lower median D$_n$4000 value compared to non-AGN galaxies (Table \ref{table_sfh_numbers}). However, the observed differences of the two D$_n$4000 distributions are not statistically significant ($\rm p-value=0.31$). 

At the highest L$_X$ range spanned by our sample, our analysis indicates that AGN have significantly lower median D$_n$4000 values compared to non-AGN galaxies (Table \ref{table_sfh_numbers}). The differences in the D$_n$4000 distributions of the two populations have also high statistical significance ($\rm p-value=0.0009$). These results suggest that luminous AGN live in galaxies with younger stellar populations compared to galaxies that do not host an active SMBH. We note, though, that our sample includes only four AGN in the highest L$_X$ range. These results are in agreement with those presented in the previous section (see also Fig. \ref{fig_sfrnorm_lx}), in the sense that, for instance, systems with higher SFRs are expected to have younger stars compared to systems with lower SFRs.

Fig. \ref{fig_sfh} and Table \ref{table_sfh_numbers} present our results for AGN and non-AGN systems in high density fields. Moderate L$_X$ AGN present lower D$_n$4000 values compared to lower L$_X$ AGN and non-AGN galaxies. Application of a KS-test reveals $\rm p-values$ of 0.02 and 0.08, between the D$_n$4000 distributions of low L$_X$ AGN and non-AGN and moderate L$_X$ AGN and non-AGN galaxies, respectively. For the most luminous X-ray sources in our sample, with $\rm log\,[L_{X,2-10keV}(erg\,s^{-1})>44$, we find that AGN have statistically significant different D$_n$4000 distributions ($\rm p-value=0.0001$) and lower D$_n$4000 values compared to non-AGN galaxies. These results corroborate the findings of the previous section.

\cite{Mountrichas2023c} used AGN in the COSMOS field and found that in low density fields, AGN with $\rm log\,[L_{X,2-10keV}(erg\,s^{-1})]<43$ tend to have higher D$_n$4000 values, while AGN with moderate L$_x$ ($\rm 43<log\,[L_{X,2-10keV}(erg\,s^{-1})]<44$) have similar D$_n$4000 values with non-AGN systems. Albeit, only the former difference appeared statistically significant at a level $>2\,\sigma$. Our results, shown in the top panel of Fig. \ref{fig_sfh} and Table \ref{table_sfh_numbers}, corroborate these previous findings. Our results are also in line with those presented in \cite{Mountrichas2023c} for AGN and non-AGN systems in high density fields.

In the study of \cite{Mountrichas2023c}, a trend was observed wherein galaxies tend to exhibit higher D$_n$4000 values as they transition to denser cosmic environments \citep[e.g.,][]{Cucciati2010, Delgado2022, Siudek2022, PerezMillan2023}. However, in our dataset, galaxies exhibit similar median D$_n$4000 values regardless of their cosmic surroundings \citep[e.g.,][]{Leslie2020, Cooke2023}. It is worth mentioning that prior research, such as \cite{Mountrichas2022c}, has highlighted the divergence in the SFH among non-AGN galaxies based on their morphological characteristics, with a comparatively lesser distinction observed among galaxies hosting AGN. Hence, the observed variations in the comparison of D$_n$4000 across diverse density fields may be, in part, influenced by the inclusion of different morphological types within the sample being examined. These morphological variations may be also the cause of the observed differences in the D$_n$4000 distributions of galaxies in different density fields. To elaborate, galaxies in low density fields have a highly peaked D$_n$4000 distribution, whereas in high density environments galaxies have a wider D$_n$4000 distribution that also presents a long tail towards high values. Although, these differences do not appear statistically significant ($\rm p-value=0.07$), similar features were also reported in the D$_n$4000 distributions of the galaxy sample used in \cite{Mountrichas2023c}. Other studies have also reported higher D$_n$4000 (and lower H$_\delta$) values for galaxies in denser fields \citep[e.g.,][]{PerezMillan2023}. \cite{Mountrichas2023c} found that this tail is populated, mainly, by quiescent galaxies that are more prominent in denser fields. \cite{Sobral2022}, found that the D$_n$4000 and H$_\delta$ indices of galaxies depend on the cosmic environment (and M$_*$), however only in the case of quiescent systems, whereas for star-forming galaxies, the two spectral indices do not show a significant dependence on environemnt. 


In summary, our analysis yields results that align with previous research, particularly the findings presented in \cite{Mountrichas2023c}, especially for AGN with low-to-moderate L$_X$. Most importantly, the results of our analysis suggest that highly luminous X-ray AGN, with $\rm log\,[L{_X,2-10keV}(erg,s^{-1})>44$, tend to reside in galaxies characterized by younger stellar populations when contrasted with their counterparts lacking an active SMBH. Furthermore, this trend appears to hold true regardless of the cosmic environment in which these sources are located. We note, though, that larger datasets of luminous AGN with available information on their local density is required to allow us to draw strong conclusions.


\section{Conclusions}
\label{sec_conclusions}

We used 149 X-ray AGN drawn from the XMM-XXL sample and 3\,488 non-AGN galaxies included in the VIPERS galaxy catalogue, to study the star-formation rates and star-formation histories of the two populations, across difference density fields. Our sources span a redshift range of $\rm 0.5<z<1.0$, have stellar masses within $\rm 10.5<log\,[M_*(M_\odot)]<11.5$ and encompass a broad spectrum of L$_X$ ($\rm 42<log\,[L_{X,2-10keV}(ergs^{-1})]<45$). To distinguish between sources located in high and low-density fields, we utilized the measurements of the overdensity parameter ($\rm log(1+\delta)$) available in the VIPERS catalog. In our analysis, we conducted SED fitting using the CIGALE code to extract various properties of the (host) galaxies. We maintained stringent photometric and quality criteria to ensure the inclusion of sources with reliable calculations in our study. In summary, our primary findings can be summarized as follows:

\begin{itemize} 

\item[$\bullet$] Low and moderate L$_X$ AGN ($\rm 42<log\,[L_{X,2-10keV}(ergs^{-1})]<44$) that live in low density fields exhibit a nearly constant SFR$_{norm}-$L$_X$ relationship. The SFR of X-ray AGN that span such luminosities appears lower, or at most comparable, to that of non-AGN galaxies.

\item[$\bullet$] AGN situated in high density environments, present an increase of SFR$_{norm}$ with L$_X$, across all L$_X$ spanned by our dataset.

\item[$\bullet$] The most luminous of the X-ray sources ($\rm log\,[L_{X,2-10keV}(ergs^{-1})]>44$ display an elevated SFR compared to non-AGN galaxies, regardless of the density field they live in.

\item[$\bullet$] For similar L$_X$, SFR$_{norm}$ appears to be higher for sources in high density regions compared to those in low density fields. While this difference falls within the statistical uncertainties associated with SFR$_{norm}$ measurements, it remains consistent across the entire L$_X$ range covered by our sources.

\item[$\bullet$] Low and moderate L$_X$ AGN live in galaxies that have comparable stellar populations (i.e., have similar D$_n$4000 values) with non-AGN galaxies, regardless of the density field they live in.

\item[$\bullet$] Luminous AGN tend to have younger stellar populations (lower D$_n$4000 values) compared to non-AGN galaxies, irrespective of the density field they reside.

\end{itemize}

Our findings may imply that in dense environments the AGN feedback acts against the removal of the gas and prevents the suppression of star formation. The most powerful the AGN, the more effective its feedback becomes. Nevertheless, another way to interpet our finding might be a scenario in which a shared mechanism, such as mergers, serves to fuel both the star formation and the SMBH. The greater the amount of gas channeled to the galaxy through the triggering mechanism, the more pronounced the increase in both the SFR and X-ray luminosity. The lower SFR observed in galaxies hosting AGN, compared to non-AGN systems (SFR$)_{norm}<1$), in low-density fields could be attributed to a scarcity of available gas in these systems, indicative of negative AGN feedback \citep{Zubovas2013}. For a SMBH to accrete at very high X-ray luminosities ($\rm log,[L_{X,2-10keV}(ergs^{-1})]>44$), an ample supply of gas is required, and this gas concurrently fuels the SFR in AGN-hosting systems, regardless of the galaxy's environment.

These results contribute to our understanding of the complex interplay between SMBHs and their host galaxies, shedding light on how environmental factors and AGN luminosity influence star formation and the stellar populations of these systems. However, it is worth emphasizing that further studies with larger datasets will be essential to refine and validate these intriguing trends and provide a more comprehensive understanding of the underlying processes at play.

\begin{acknowledgements}
This project has received funding from the European Union's Horizon 2020 research and innovation program under grant agreement no. 101004168, the XMM2ATHENA project. This work has been supported by the Polish National Agency for Academic Exchange (Bekker grant BPN/BEK/2021/1/00298/DEC/1), the European Union's Horizon 2020 Research and Innovation programme under the Maria Sklodowska-Curie grant agreement (No. 754510).

\end{acknowledgements}

\bibliography{mybib}
\bibliographystyle{aa}

\end{document}